\documentclass[twocolumn]{pasj00}

\Received{2012 March 12}
\Accepted{2012 April 15}
\SetRunningHead{N. Sakai et al.}{Astrometry of IRAS 05168+3634}

\usepackage{times}
\usepackage{supertabular}
\usepackage{subfigure}

\begin{document}

\title{Outer Rotation Curve of the Galaxy with VERA I:\\ Trigonometric parallax of IRAS 05168+3634}
\author{Nobuyuki Sakai$^{1}$, Mareki Honma$^{1,2}$, Hiroyuki Nakanishi$^{3}$, Hirofumi Sakanoue$^{3}$, Tomoharu Kurayama$^{3}$, Katsunori M. Shibata$^{1,2}$, and Makoto Shizugami$^{2}$%
}
\affil{%
   $^{1}$The Graduate University for Advanced Studies (Sokendai), Mitaka, Tokyo 181-8588}
\affil{%
$^{2}$Mizusawa VLBI Observatory, National Astronomical Observatory of Japan, Mitaka, Tokyo 181-8588}
\affil{%
$^{3}$Faculty of Science, Kagoshima University, 1-21-35 Korimoto, Kagoshima, Kagoshima 890-0065}
\email{nobuyuki.sakai@nao.ac.jp}

\maketitle

\begin{abstract}
 We report measurement of trigonometric parallax of IRAS 05168+3634 with VERA. The parallax is 0.532 $\pm$ 0.053 mas, corresponding to a distance of 1.88$^{+0.21}_{-0.17}$ kpc.
This result is significantly smaller than the previous distance estimate of 6 kpc based on kinematic distance. 
This drastic change in the source distance revises not only physical parameters of IRAS 05168+3634, but also its location of the source, placing it in the Perseus arm rather than the Outer arm.
 We also measure  proper motions of the source. A combination of the distance and the proper motions with  systemic velocity yields rotation velocity ($\Theta$) of 227$^{+9}_{-11}$ km s$^{-1}$ at the source, assuming $\Theta_{\rm{0}}$ = 240 km s$^{-1}$.
Our result combined with previous VLBI results for six sources in the Perseus arm indicates that the sources rotate systematically slower than the Galactic rotation velocity at the LSR.
 In fact, we show  observed disk peculiar motions averaged over the seven sources in the Perseus arm as ($U_{\rm{mean}}$, $V_{\rm{mean}}$) = (11 $\pm$ 3, $-$17 $\pm$ 3) km s$^{-1}$, 
indicating that these seven sources are systematically moving toward the Galactic center, and lag behind the Galactic rotation.
\end{abstract}

\section{Introduction}
 The rotation curve can be used to determine mass distribution in spiral galaxies through equilibrium between the gravity and the centrifugal force. 
It has been revealed that there exist plenty of flat rotation curves beyond optical disks in external spiral galaxies. 
This indicates the existence of large quantities of dark matter in the outer regions of galaxies (e.g. Sofue et al. 1999).
In contrast, the rotation curve of the Milky Way has  a relatively large ambiguity,  although it is believed to be almost flat, being similar to the external spiral galaxies (e.g. Sofue et al. 2009). 
This uncertainty is mainly due to difficulties in measuring distance, since we are within the Milky Way. 
To precisely measure  the  distance and proper motions of Galactic objects based on astrometry, the HIPPARCOS satellite was launched in 1989 (Perryman, 1989). 
However, parallax measurement with HIPPARCOS was limited to within 100 pc from the solar position, in contrast to the Milky Way's disk size of $\sim$ 20 kpc. 
Today, well-developed interferometer techniques in radio wave length can be used to conduct Galactic astrometry over a kilo-parsec scale. 
In fact, VERA (VLBI Exploration of Radio Astrometry) and VLBA (Very Long Baseline Array) have succeeded in  making  distance measurements over 5 kpc (e.g. Nagayama et al. 2011b, Sanna et al. 2012). 
To construct a rotation curve of the Milky Way in the outer region with high accuracy, we have been using VERA to observe Galactic objects with H$_{2}$O maser emissions in the Galactic-outer region. 
In this paper, we report the observation results for IRAS 05168+3634.

\begin{figure*}[t]
  \begin{center}
\FigureFile(160mm,139mm){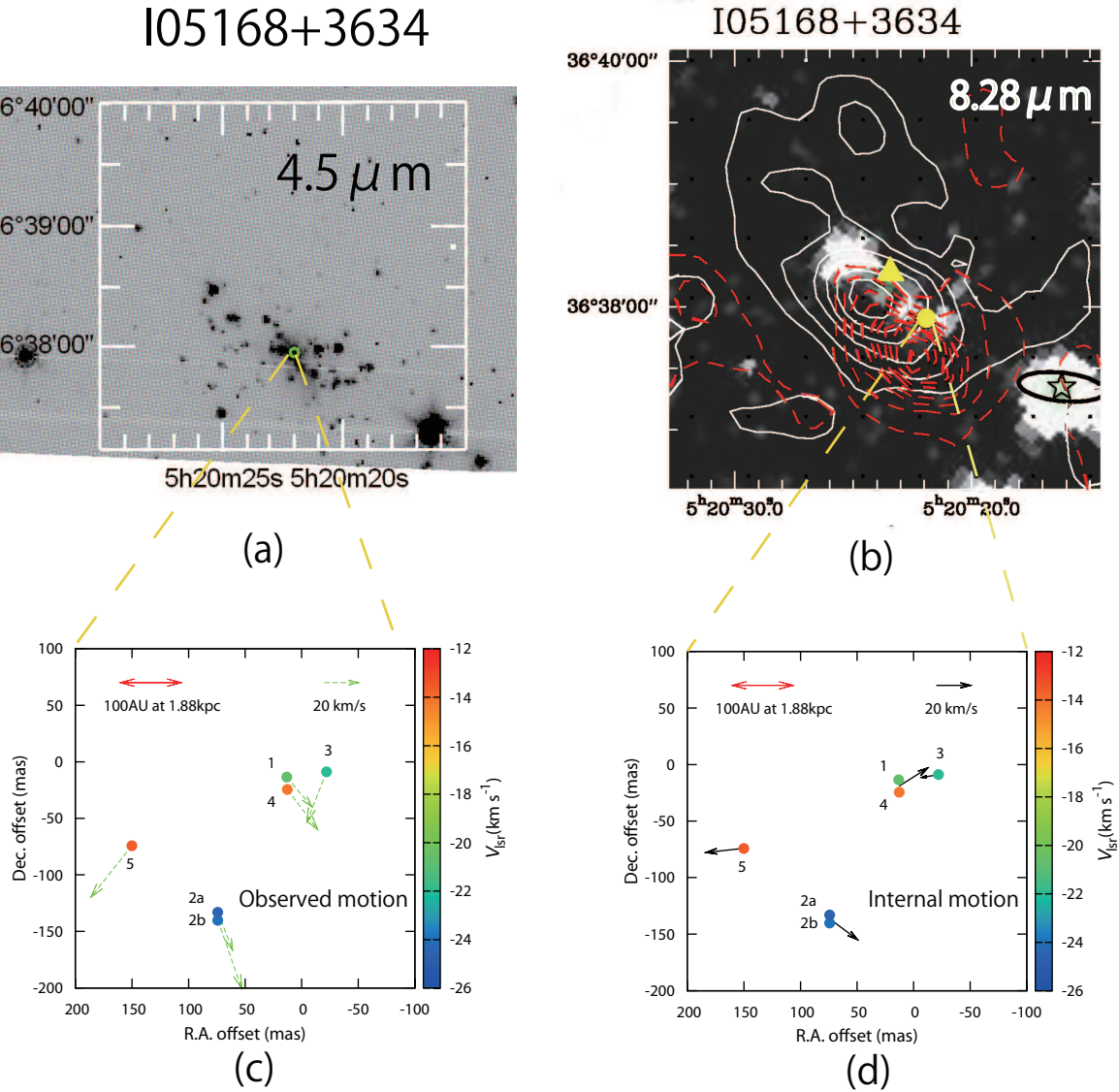}
 \end{center}
    \caption{ 
(a) Spitzer IRAC image at 4.5$\mu$m taken from NASA/IPAC Infrared Science Archive with the circle corresponding to the observed maser position at ($\alpha_{\rm{J2000}}$, $\delta_{\rm{J2000}}$) = (\timeform{5h20m22.070s}, \timeform{36d37'56".63}). 
(b) The circle representing the observed maser position is superposed on the CO ($J$ = 2-1) contours map (Zhang et al. 2005). The dashed contours denote the red-shifted emission, and the solid contours represent the blue-shifted emission.
 Masers are associated with CO ($J$ = 2-1) outflow. The gray scale shows an MSX A-band image (8.28$\mu$m) taken from Price et al. (2001). 
The symbol of ($\ast$) and the associated ellipse mark the position and the position uncertainty of the IRAS source. The triangle marks the position of UC HII region taken from Molinari et al. (1998).
(c) Maser distribution map. The nominal origin is set to the phase tracking center. The green vectors represent direct observed motions with respect to the position reference source. (d)Same as (c) but with the internal motions. The black vectors show the internal motions with the systematic motions subtracted (see  text). }
    \label{fig.1}
\end{figure*}

\begin{table*}[t]
\begin{center}
\small
\caption{Observations}
\begin{tabular}{lcccc}
\hline
\hline
Epoch	&Date	&Time Range	&Beam	&Note	\\ 
  		&	&(UTC)		&(mas)	&	\\
\hline
  A	&2009 Oct 03&15:35-00:40 &1.15$\times$ 0.72@139$\fdg$0 & \\
  B	&2009 Oct 30&13:50-22:15 &1.16$\times$ 0.72@135$\fdg$0 & \\
  C	&2010 Jan 25&08:05-16:30 &1.04$\times$ 0.69@131$\fdg$5 & \\
  D	&2010 Mar 21&04:30-12:55 &1.15$\times$ 0.70@133$\fdg$5 &Two maser sources were alternately observed in the same epoch.\\
  E	&2010 May 22&00:25-08:50 &1.21$\times$ 0.68@149$\fdg$1 &			//	\\
  F	&2010 Aug 22&18:20-02:45 &1.12$\times$ 0.72@132$\fdg$7 &			//	\\
  G	&2010 Oct 24&14:15-22:40 &1.09$\times$ 0.73@135$\fdg$2 &			//	\\
  H	&2010 Dec 02&11:40-20:05 &1.18$\times$ 0.69@131$\fdg$2 &			//	\\
  I	&2011 Jan 28&07:55-16:20 &1.10$\times$ 0.74@132$\fdg$9 &			//	\\
  J	&2011 Apr 11&03:10-11:35 &1.39$\times$ 0.76@168$\fdg$6 &			Mizusawa station did not participate due to system trouble.	\\
  K	&2011 May 06&01:30-11:35 &1.14$\times$ 0.66@132$\fdg$6 &			Two maser sources were alternately observed in the same epoch.	\\

\hline
\multicolumn{0}{@{}l@{}}{\hbox to 0pt{\parbox{180mm}{\footnotesize
\par\noindent
}\hss}}
\end{tabular}
\end{center}
\end{table*}

\begin{table*}[th]
\caption{Source Data}
\begin{tabular}{lccccc}
\hline
\hline
Source Name	&R.A.	&Decl.	&S.A.\footnotemark[$*$]	&Flux Density	&Note	\\
		&(J2000.0)	&(J2000.0)				&($^\circ$)			&(Jy)	&\\
\hline
IRAS 05168+3634	&\timeform{05h20m22.07s}	&\timeform{36d37'56".63}	&	&10.0 $\sim$255.6	&H$_{2}$O masers	\\
J0530+3723		&\quad \timeform{05h30m12.5493s}\footnotemark[$\dag$]        &\timeform{37d23'32".6197}\footnotemark[$\dag$]	&2.128	&0.10$\sim$0.18	&Phase-reference calibrator	\\
\hline
\multicolumn{4}{@{}l@{}}{\hbox to 0pt{\parbox{180mm}{\footnotesize
\par\noindent
\footnotemark[$*$]The separation angle between the maser and the reference sources.\\
\footnotemark[$\dag$]The position is based on Fomalont et al. (2003).
}\hss}}
\end{tabular}
\end{table*}

IRAS 05168+3634 is a high-mass star-forming region in the pre-UC HII phase (Wang et al. 2009). 
According to Wang et al. (2009), the source has a $^{13}$CO ($J$ = 2-1) core which is positioned $\sim$ 1.3' toward the northeast of  IRAS 05168+3634 and has an active outflow. 
An H$_{2}$O maser is detected in this region where the CO outflow occurs (Palla et al. 1991). 
The maser position coincides with a bright red source detected with the Spitzer IRAC at 4.5 $\mu$m (fig. 1a). 
Figure 1b represents the MSX A-band image (8.28$\mu$m, Price et al. 2001) on which  CO ($J$ = 2-1) emissions (Zhang et al. 2005), a 6 cm continuum emission peak (Molinari et al. 1998), and the IRAS source position are superposed . 
Although no radio continuum (6cm) was detected in IRAS 05168+3634 by McCutcheon et al. (1991), Molinari et al. (1998) have detected the  6cm continuum emission peak with 4.23 mJy at the triangle position in figure 1b (Zhang et al. 2005). 
In this region, there are several sources with different evolutionary phases of star-formation. 
Molinari et al. (1998) interpreted the lack of radio emission in this star-forming region as being due to accreting matter that chokes off expansion of ionized gas.
As for the distance-estimate of IRAS 05168+3634, kinematic distance of $\sim$6 kpc was obtained (Molinari et al. 1996) based on  the systemic LSR velocity ($V_{\rm{LSR}}$) of $-$15.5 $\pm$ 1.9 km s$^{-1}$ traced by CS (2-1) emission (Bronfman et al. 1996). 
However, this source is located at the Galactic longitude ($\ell$) of $\sim$ 171$^{\circ}$ where kinematic distance has a large error (Sofue et al. 2011a).
In fact, kinematic distances along the Sun-Galactic Center (G.C.) line are not determined precisely since  the $V_{\rm{LSR}}$ is degenerated around 0 km s$^{-1}$. 
To estimate error of the kinematic distance at the source, we consider the $\Delta V_{\rm{LSR}}=\pm$1.9 km s$^{-1}$ with the flat rotation model. The obtained error is significantly large, $\sim$ 1.2 kpc or $\sim$22$\%$ at the source. Thus, VLBI observations for IRAS 05168+3634 are necessary, 
not only to precisely measure distance, but also to construct a precise Outer Rotation Curve of the Galaxy, 
and in the present paper we report the astrometric observations of IRAS 05168+3634 with VERA.

\section{Observations and Data Reduction}
\subsection{VLBI Observations with VERA}
 Between October 2009 and May 2011, we carried out 11-epoch observations of H$_{2}$O maser line at  a rest frequency of 22.235080 GHz to measure parallax and proper motions of IRAS 05168+3634. Details of dates for the 11 observations are listed in table 1.
  Note that system troubles prevented Mizusawa station from participating in observation on DOY (day of year) 101 of 2011. 
Except for  the observation, average synthesized-beam was 1.1 $\times$ 0.7 mas with a position angle of 135$^{\circ}$ as listed in table 1. In contrast, the synthesized beam was 1.4 $\times$ 0.8 mas with a position angle of 169$^{\circ}$ in the observation on DOY 101, 2011. 
We also observed two pairs, both including a target source and a reference source from DOY 80, 2010 for the phase referencing observation.
One of these is target source IRAS 05168+3634 and reference source J0530+3723.
The other is target source IRAS 04579+4703 and reference source J0507+4645. 
Both a maser source and a QSO were observed simultaneously using the dual-beam mode (Kobayashi et al. 2008). 
Observational results for the latter source pair will be reported in another paper. 
Fringe finder source J0555+3948 or DA193 was also observed and used for calibration of clock parameters in correlation processing. 
On-source time for IRAS 05168+3634 was 3.1 hours in every observation from DOY 80, 2010, compared with a total observation time of about eight hours. 
In contrast, on-source time was 5.6 hours between DOY 276 in 2009 and DOY 25 in 2010 for the former three observations. 
During 11 observations, we detected maser emissions in all epochs for IRAS 05168, and found that maser flux varied significantly between $\sim$10 Jy and $\sim$200 Jy in the observations (table 2). 
Table 2 also shows variation of flux for J0530+3723, the tracking centers of the maser and the reference sources, and a separation angle between the two sources. 
During these observations, left-hand circular polarization was recorded onto magnetic tapes at a rate of 1024 Mbps with 2-bit quantization, after filtering was performed by the VERA digital filter. 
Total band width of 256 MHz consisted of 16 of 16-MHz IF sub-bands. One of the 16-IFs was assigned for the maser source, and the other 15-IFs were assigned for the continuum sources as reference and fringe finder sources.
Magnetic tapes in four stations were delivered to NAOJ to conduct correlation processing with the Mitaka FX correlator. 
The correlator accumulation period was 1 sec. The one IF was composed of 512 channels for the maser source, leading to a frequency resolution of 31.25 kHz and a velocity resolution of 0.42 km s$^{-1}$. 
In contrast, each 15-IFs was composed of 64 channels for the continuum sources.

\begin{table*}[t]
\small
\begin{center}
\caption{Parallax Fits\footnotemark[$*$]}
\begin{tabular}{ccccccccccc}
\hline
\hline
Feature	&$V_{\rm{LSR}}$	&N$_{\rm{epochs}}$	&Epochs	&Parallax(Error)	&\multicolumn{2}{c}{Proper Motions(Error)}	&	&\multicolumn{3}{c}{Errors	}\\
\cline{6-7} \cline{9-11}
		&km s$^{-1}$	&				&			&(mas)	&$\mu_{\alpha}$cos$\delta$	&$\mu_{\delta}$	&	&R.A.	&Dec.	&Both	\\
		&	&				&			&	&(mas yr$^{-1}$)&(mas yr$^{-1}$)&	&	&(mas)	&	\\
\hline
1	&$-$20.3	&9	&ABDEFHIJK	&0.548(0.104)		&$-$1.65(0.14)	&$-$1.98(0.21)	&	&0.18	&0.34	&0.38	\\
	&$-$20.7	&10	&ABCDEFHIJK	&0.533(0.087)		&$-$1.66(0.11)	&$-$1.94(0.19)	&	&0.17	&0.33	&0.37	\\
	&$-$21.1	&9	&ABDEFHIJK	&0.561(0.109)		&$-$1.62(0.14)	&$-$1.97(0.20)	&	&0.19	&0.34	&0.39	\\
2b	&$-$23.6	&9	&ABCDEFGHI	&0.538(0.080)		&$-$1.59(0.13)	&$-$4.67(0.17)	&	&0.17	&0.23	&0.29	\\
	&$-$24.1	&9	&ACDEFGHIJ	&0.519(0.066)		&$-$1.57(0.11)	&$-$4.41(0.23)	&	&0.15	&0.30	&0.34	\\
\hline
\multicolumn{2}{l}{Combined fit for five spots}	&	&	&0.537(0.038)	&	&	&	&0.17	&0.31	&0.35	\\
\multicolumn{2}{l}{Combined fit for two features}	&	&	&0.526(0.053)	&	&	&	&0.16	&0.32	&0.36	\\
\hline
\multicolumn{3}{l}{Final (mean of the two combined fittings)}		&	&0.532(0.053)	&	&	&	&	&	&	\\

\hline
\multicolumn{4}{@{}l@{}}{\hbox to 0pt{\parbox{180mm}{\footnotesize
\par\noindent
\\
\footnotemark[$*$]Combined fits were done to the data set of two (see text). Final value is determined by taking the mean of the two.\\ 
Error of the final parallax is estimated by combining in quadrature the scatter of the individual parallaxes around the mean ($\pm$0.006 mas)\\ and the error bar of individual parallax
 ($\pm$0.053 mas).
}\hss}}
\end{tabular}
\end{center}
\end{table*}

\begin{table*}
\begin{center}
\caption{Determination of the systematic proper motions for IRAS 05168+3634}
\begin{tabular}{ccccc}
\hline
\hline
Feature	&	&\multicolumn{2}{c}{Proper Motions\footnotemark[$*$] (Error)}	&Note\\
\cline{3-4}
	&$V_{\rm{LSR}}$	&$\mu_{\alpha}$cos$\delta$	&$\mu_{\delta}$	&\\
	&km s$^{-1}$	&(mas yr$^{-1}$)&(mas yr$^{-1}$)	&\\
\hline
1 $\&$ 4	&$-$20.3 $\sim$ $-$21.1 $\&$ $-$13.6 $\sim$ $-$14.8	&$-$1.83(0.25)	&$-$2.31(0.49)	&		\\
2a $\&$ 2b	&$-$24.5 $\&$ $-$23.6 $\sim$ $-$24.1			&$-$1.28(0.43)	&$-$3.52(1.43)	&		\\
3	&$-$21.1 $\sim$ $-$22.8	&1.30(0.24)		&$-$3.31(0.18)	&		\\
5	&$-$13.6	&2.71(0.81)	&$-$3.42(1.62)	&		\\

\hline
Mean	&$-$19.3	&0.23(1.07)\footnotemark[$\dag$]	&$-$3.14(0.28)\footnotemark[$\dag$]	&Systematic proper motions\\

\hline
\multicolumn{4}{@{}l@{}}{\hbox to 0pt{\parbox{180mm}{\footnotesize
\par\noindent \\
\footnotemark[$*$]We averaged the proper motion values of four representative features to determine the systematic proper motions \\(see text).
Note that these four feature values were determined by adapting the parallax of 0.532 mas.  \\
\footnotemark[$\dag$]The errors of the systematic proper motions were determined by dividing the standard deviations by a factor of \ $\sqrt[]{4}$.\\
\\
}\hss}}

\end{tabular}
\end{center}
\end{table*}

\subsection{Data Reduction}
 The Astronomical Image Processing System (AIPS, NRAO) was used for data calibration, and general phase referencing analysis was applied to determine absolute maser positions (e.g. see Appendix in Kurayama et al. 2011 for the phase referencing analysis with VERA).
 Before fringe search was done in AIPS to determine the clock offset and the clock rate offset, we corrected the delay model which was used in the correlation processing, since the delay model used in the correlation was not accurate enough for precise astrometry.
 To correct the delay model, we applied a precise geodetic model, the most-updated Earth-rotation parameters provided by IERS, tropospheric delays measured with GPS receivers at each VERA station (Honma et al. 2008a),
 and ionospheric delays based on the Global Ionosphere Map (GIM) which was produced every two hours by the University of Bern. 
At the beginning of fringe searches with the corrected visibilities, fringes were searched for the fringe finder J0555+3948 or DA193 to determine the clock offset. 
Next, fringes were searched for the position reference J0530+3723 by referring to the clock offset determined with J0555+3948 or DA193. 
Then, the position reference was imaged with self calibration, and complex gain solved for the position reference was transferred to the maser source IRAS 05168+3634 for the phase-referencing. 
When transferring the complex gain from J0530+3723 to IRAS 05168+3634, the instrumental phase difference between the dual-beam system with VERA was corrected based on the phase difference between the dual-beam system measured in real time during  the  observations (Honma et al. 2008b).
 At the next step, the corrected visibilities for the maser source were Fourier-transformed to create dirty images of each maser channel. 
Finally, to obtain final CLEANed maps, we used the ``clean``  task in AIPS and we identified masers at each velocity channel from the final maps. 
We used the ``jmfit`` task in AIPS to determine an absolute maser position and a flux of the maser source by elliptical Gaussians fitting to the brightness peak of the map. 
We regarded a maser detected  as it achieved a high SNR (signal to noise ratio) of five or more. 
To identify the same maser spot in each epoch for astrometry, we used the following selection criteria: 
1) spots should be in the same velocity channel and 2) the difference in the positions of spots between epochs should not exceed the proper motion threshold, which was set to be 10 mas yr$^{-1}$. 
Among the maser spots identified using these criteria, spots detected in three epochs or more were used for proper motion determinations,  while spots detected in nine epochs or more with two or more continuously channels were used for parallax determinations. 
As a result of data analyses with the above criteria, we detected six maser features with additional maser spots to determine proper motions, and two maser features were detected to determine a parallax (see fig. 1c, fig. 1d, table 3, and table 4). 
In  a  model fitting process for the parallax and the proper motion determinations, we used ``VERA$\underline{\hspace{0.5em}}$Parallax``, one of the tasks performed by the ``VEDA (VEra Data Analyzer)``, data analyzing software developed at NAOJ.
For astrometry, we assumed that source motions can be described by a combination of linear proper motion and sinusoidal parallax motion. 

\begin{figure*}[t]
\FigureFile(80mm,59.65mm){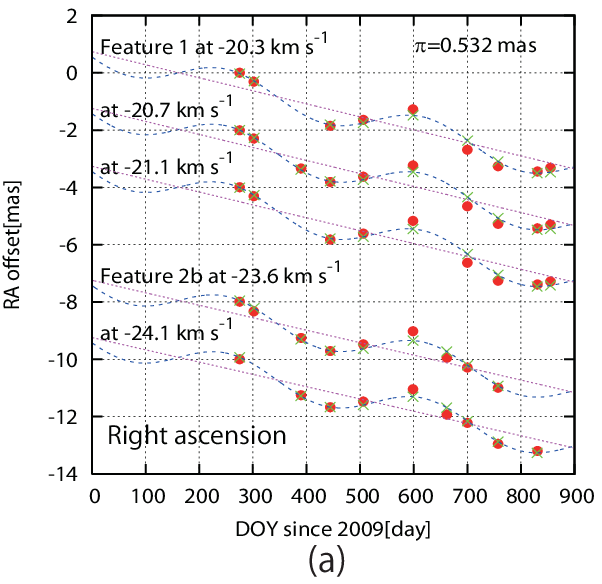}%
\FigureFile(80mm,59.65mm){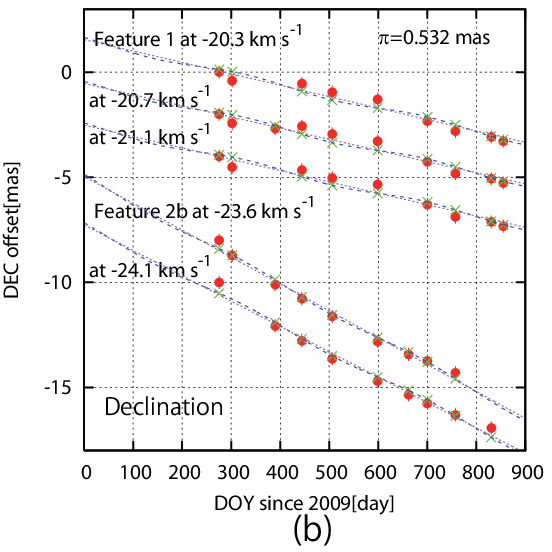}\\
\FigureFile(80mm,59.65mm){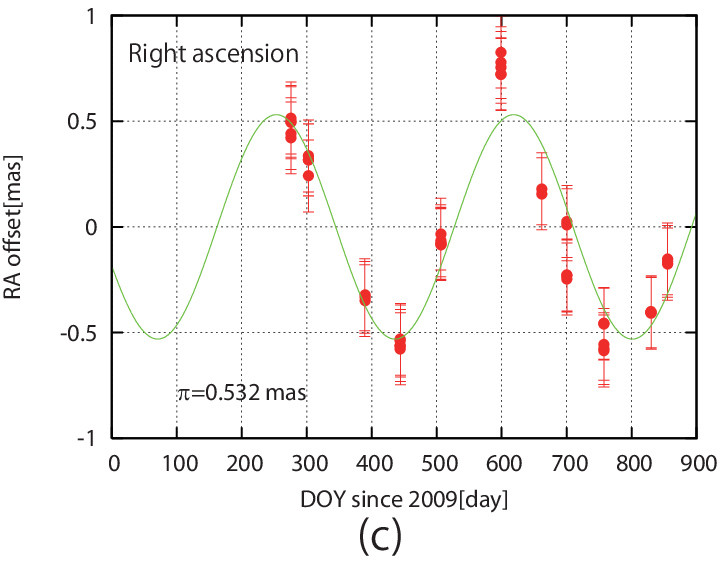}%
\FigureFile(80mm,59.65mm){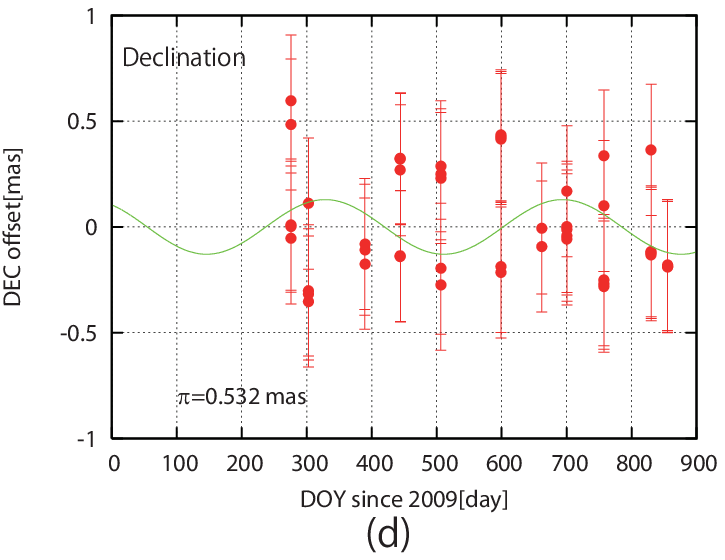}\\
\caption{Maser positional evolutions and the combined-fit results (see text). The error bars represent position errors resulting from the astrometric (systematic) errors which are given so that the reduced $\chi^{2}$ becomes unity.
(a) Maser positional evolutions in the right ascension with the circles. Dotted lines represent proper motions, and dashed lines show fitted lines. (b) Same as (a) but in the declination.
 (c) Parallax motions in the right ascension with the proper motions subtracted from (a). The solid lines show a line fitted to the parallax motions with the circles. (d) Same as (c) but in the declination.
}
\end{figure*}

\section{Results}
\subsection{Trigonometric parallax of IRAS 05168+3634}
  The  11-epoch VLBI observations  performed over about one and a half years clearly show  an  annual parallax and proper motions for IRAS 05168+3634 (see table 3 and fig. 2). 
In table 3, we show the  parallax results for five spots in the two features (1 and 2b) to be 0.548 $\pm$ 0.104, 0.533 $\pm$ 0.087, 0.561 $\pm$ 0.109, 0.538 $\pm$ 0.080, and 0.519 $\pm$ 0.066 mas, respectively. Note that astrometric errors in each epoch are given so that the  reduced  $\chi^{2}$ becomes unity since  a  systematic error is generally larger than  a  thermal error estimate in the phase-referencing VLBI (e.g. Sanna et al. 2012). 
In addition, we conducted combined-fitting assuming a common parallax with discrete proper motions for each spot. 
As a result, the parallax was determined to be 0.537 $\pm$ 0.038 mas for  the  five spots, corresponding to a distance of 1.86$^{+0.14}_{-0.12}$ kpc.
In the same way, the other parallax was determined to be 0.526 $\pm$ 0.053 mas for  the  two features (feature-1 with $V_{\rm{LSR}}$ = $-$20.7 km s$^{-1}$ and feature-2 with $V_{\rm{LSR}}$ = $-$24.1 km s$^{-1}$), corresponding to a distance of 1.90$^{+0.21}_{-0.17}$ kpc.
Note that the selected two spots in the two features are brighter than the other spots and have less parallax errors compared with the other spots.
We averaged the two combined fit results to obtain a final parallax result since the five spots may not be independent of each other.
As a consequence, we obtained a final parallax of 0.532 $\pm$ 0.053 mas, corresponding to a distance of 1.88$^{+0.21}_{-0.17}$ kpc. 
Note that the parallax error was estimated by combining in quadrature the scatter of the individual parallaxes around the mean ($\pm$0.006 mas) and the error bar of individual parallax ($\pm$0.053 mas).
In figures 2a and 2b, we show the combined fit results with the final parallax result fixed for the five spots in directions of right-ascension (R.A.) and declination (Dec.), respectively.
 The  error bars in figure 2 represent the astrometric error described above. 
 The  astrometric errors set for the combined fitting are 0.17 mas for $\Delta \alpha \cos\delta$ and 0.31
mas for $\Delta \delta$. These errors mainly originated in the tropospheric zenith delay (e.g. Honma et al. 2007).
Moreover, the separation angle between the maser and the phase reference sources is relatively large (= 2$\fdg$1) in our observations, which causes a large residual of tropospheric delay between the target and the reference pair. 
The error of $\Delta \delta$ is larger than that of $\Delta \alpha \cos\delta$, which is consistent with previous VERA results. In addition, the parallax amplitude is reduced in the declination. Hence, the parallax signal is not clear in the declination as fig. 2d. 
\begin{figure*}[tbp]
\FigureFile(75.0mm,77.5mm){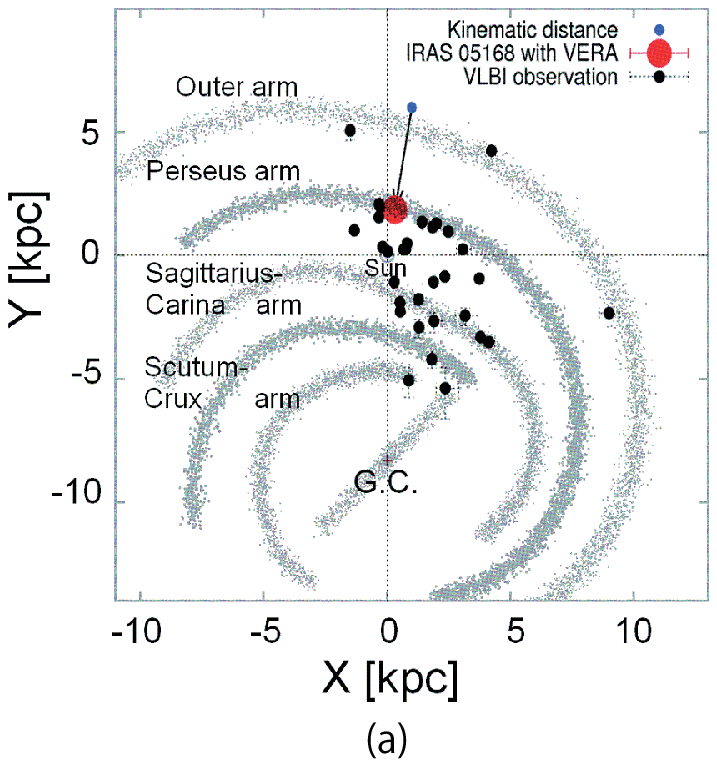}%
\FigureFile(85.0mm,77.5mm){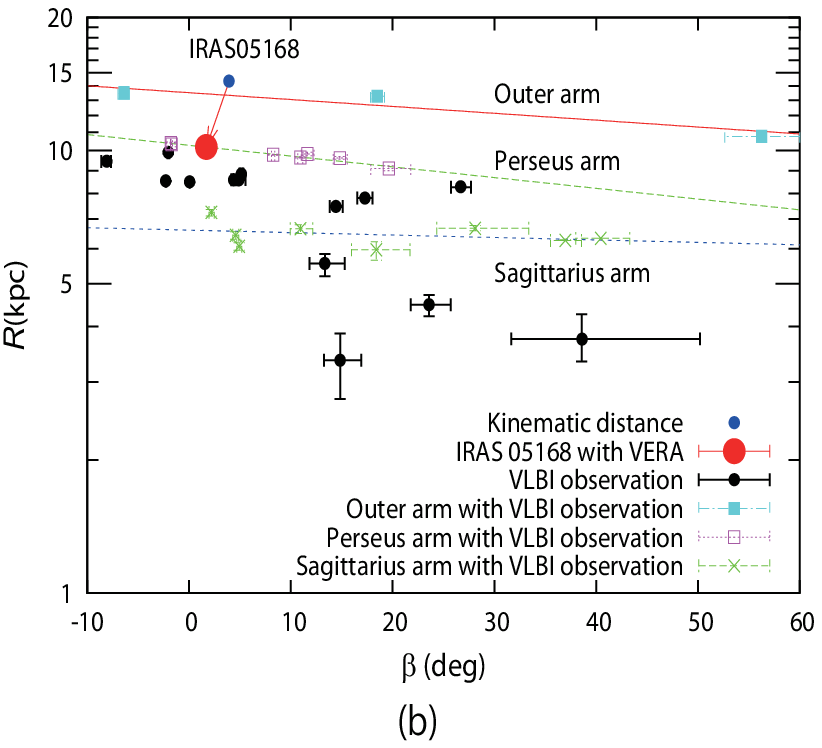}\\
\caption{The position of IRAS 05168+3634 based on both kinematic distance and our parallax measurement. (a) The red circle shows our observation result, while the blue one represents the kinematic distance. The black circles show previous VLBI results for other sources in star-forming regions (see table 6). 
These results are superposed on the Galactic face-on image (Georgelin $\&$ Georgelin 1976). 
$R_{0}$ = 8.33 kpc is assumed (Gillessen et al. 2009). The solar position is (X, Y) = (0, 0) kpc. (b) It is the same as (a) but in the polar coordinate. The horizontal axis is the Galacto-centric longitude $\beta$, and the vertical one is the Galacto-centric distance in log scale (see text). 
Solid and dashed lines show fitted lines for Outer and Perseus arms through the equation of log-spiral (see Eq. 1). 
The dotted line also represents the fitted line for the Sagittarius-Carina arm. 
The pitch angles of Outer and Perseus arms show 11$\fdg$6 and 17$\fdg$8 $\pm$ 1$\fdg$7, respectively. 
The pitch angle of Sagittarius-Carina arm is not determined precisely.}
\end{figure*}

The final parallax result of 1.88$^{+0.21}_{-0.17}$ kpc is smaller than the previously estimated source distance of 6 kpc based on kinematic distance (Molinari et al. 1996).
We will discuss the consequence of this difference in distance in chapter 4.

\subsection{Systematic proper motions of IRAS 05168+3634}
As the next step, we determine systematic proper motions from the observational data. 
Note that a maser source has both internal and systematic proper motions. Hence, to obtain the systematic motions, one should remove the internal motions (e.g. Hachisuka et al. 2009). 
Figures 1c and  1d  show distribution of maser spots that were detected for more than two epochs. 
The brightest maser spot with $V_{\rm{LSR}}$ = $-$20.7 km s$^{-1}$ in the feature-1 is located close to the nominal origin of the map, which was set to the phase-tracking center. 
Green vectors in fig. 1c represent direct observed motions with respect to the position reference source. 
These indicate that there is no clear sign of bi-polar outflow although Sato et al. (2010b) showed  the  bi-polar outflow with which H$_{2}$O masers were associated in another massive star-forming region.
Thus, here we assume random internal motions to determine the systematic proper motions. 
We identify six features to determine the systematic proper motions through the described criteria (fig. 1c. fig. 1d, and table 4). 
Features 1 and 4 are located in the same region with the same directed motion, meaning that these may be associated with the same gas. In the same way, features 2a and 2b may also be associated with the same gas. 
Thus, we reduced the number of data from six to a representative four, including features 1 $\&$ 4, 2a $\&$ 2b, 3, and 5 in table 4. 
The proper motions of the four representative features were derived by adapting  the parallax of 0.532 mas. 
As a result of data averaging, we determine the systematic proper motions of ($\mu_{\alpha} \mathrm{cos}\delta$,\ $\mu_{\delta}$) = (0.23 $\pm$ 1.07,\ $-$3.14 $\pm$ 0.28) mas yr$^{-1}$ in the equatorial coordinate. 
Note that the errors of proper motions were determined by dividing the standard deviations of proper motions by a factor of $\sqrt{4}$. 
Black vectors in figure 1d represent  the  internal motions with the systematic motions subtracted, which in fact appears basically random.
However, the error is relatively large for the systematic proper motions, since only four features were used to determine the systematic proper motions. 
To express the large error, we compared the difference between the LSR velocity of $-$15.5 $\pm$ 1.9 km s$^{-1}$ for IRAS 05168+3634 (Bronfman et al. 1996) and the averaged $V_{\rm{LSR}}$ of $-$19.3 km s$^{-1}$ for the four representative features.
The difference of 3.8 km s$^{-1}$ is converted to 0.43 mas yr$^{-1}$ by adapting the distance of 1.88 kpc for IRAS 05168.
 This indicates that the determined systematic proper motions for IRAS 05168 include an error of at least 0.4 mas yr$^{-1}$.

\begin{table*}[ht]
\begin{center}
\caption{Revised physical parameters for IRAS 05168+3634}
\begin{tabular}{ccc}
\hline
\hline
Physical parameter	&Kinematic distance of 6.08 kpc (Molinari et al. 1996)	&Our parallax measurement of 1.88 kpc\\

\hline
Virial mass ($M_{\odot}$)		&2.4$\times$10$^{3}$			&7.4$\times$10$^{2}$			\\
LTE mass ($M_{\odot}$)		&$>$1.2$\times$10$^{4}$			&$>$1.1$\times$10$^{3}$			\\
$\alpha$=$M_{\rm{vir}}$ / $M_{\rm{LTE}}$	&0.2				&0.7				\\
Bolometric luminosity ($L_{\odot}$)		&17,130			&1638				\\
Spectral type 			&B0.5$^{\ast}$		&B3$^{\ast}$		\\

\hline
\multicolumn{3}{@{}l@{}}{\hbox to 0pt{\parbox{180mm}{\footnotesize
\par\noindent \\
\footnotemark[$*$]Panagia, 1973.
\\
}\hss}}

\end{tabular}
\end{center}
\end{table*}

\section{Discussion}

\subsection{Location and revised physical parameters for IRAS 05168+3634}\label{sec:math-symbols}
 Our parallax measurement revises the distance to IRAS 05168+3634 from 6 kpc to 1.88 kpc, which requires a significant reduction of physical parameters for the source. 
By adapting the distance of 1.88 kpc, the source is likely to be located in the Perseus arm rather than in the Outer arm (fig. 3).
  Figure 3a provides a schematic face-on view of the Galaxy (Georgelin $\&$ Georgelin 1976) on which the location of IRAS 05168+3634 is superposed. 
While the blue circle is based on the kinematic distance of the source, the red circle is based on our parallax measurement. 
Clearly, our result is smaller than the kinematic distance. 
Black circles show previous VLBI results listed in table 6. 
Note that $R_{0}$ = 8.33 kpc was assumed from Gillessen et al. (2009). Figure 3b represents the source positions on a Galacto-centric longitude versus a Galacto-centric radius plot. 
The horizontal axis is the Galacto-centric longitude $\beta$, which is set to 0 toward the Sun and increases with the direction of Galactic rotation. 
The vertical axis is the Galacto-centric radius in log scale.
Solid, dashed, and dotted lines fitted to VLBI observation results represent the logarithmic spiral described by the equation below (Reid et al. 2009b).
\begin{equation}
\mathrm{ln}(R/R_{\rm{ref}})=-(\beta-\beta_{\rm{ref}})\mathrm{tan}\psi
\end{equation}
Here $R$ is the Galacto-centric radius, $\beta$ is the Galacto-centric longitude, $R_{\rm{ref}}$ is the reference radius for an arm, $\beta_{\rm{ref}}$ is the reference longitude for an arm, and $\psi$ is the pitch angle. 
Eight sources in the Perseus arm are listed in table 6, and plotted on figure 3b. 
They are distributed between Galactic longitude $\ell$ = 94$\fdg$6 and 189$\fdg$0 (between Galacto-centric longitude $\beta$ = $-$1$\fdg$8 and 19$\fdg$6). 
NGC 281 in the Perseus arm is affected by the super bubble motion (Sato et al. 2008), and so this source is not used for a fitting with the equation. 
A pitch angle of 17$\fdg$8 $\pm$ 1.7 for the Perseus arm is determined by the least-square fitting, which is consistent with the previous result 16$\fdg$5 $\pm$ 3$\fdg$1 in Reid et al. (2009b) within error. 
The fitted line is shown in figure 3b with the dashed line. 
In the same way, we determine a pitch angle of 11.6$^{\circ}$ between Galactic longitude $\ell$ = 75$\fdg$3 and 196$\fdg$5 (between Galacto-centric longitude $\beta$ = $-$6$\fdg$4 and 56$\fdg$3) for the Outer arm, shown in figure 3b with the solid line. 
The pitch angle of the Sagittarius-Carina arm is not determined well from the least-square fitting.

\begin{figure*}[htbp]
\begin{center}
\FigureFile(160.0mm,113.3mm){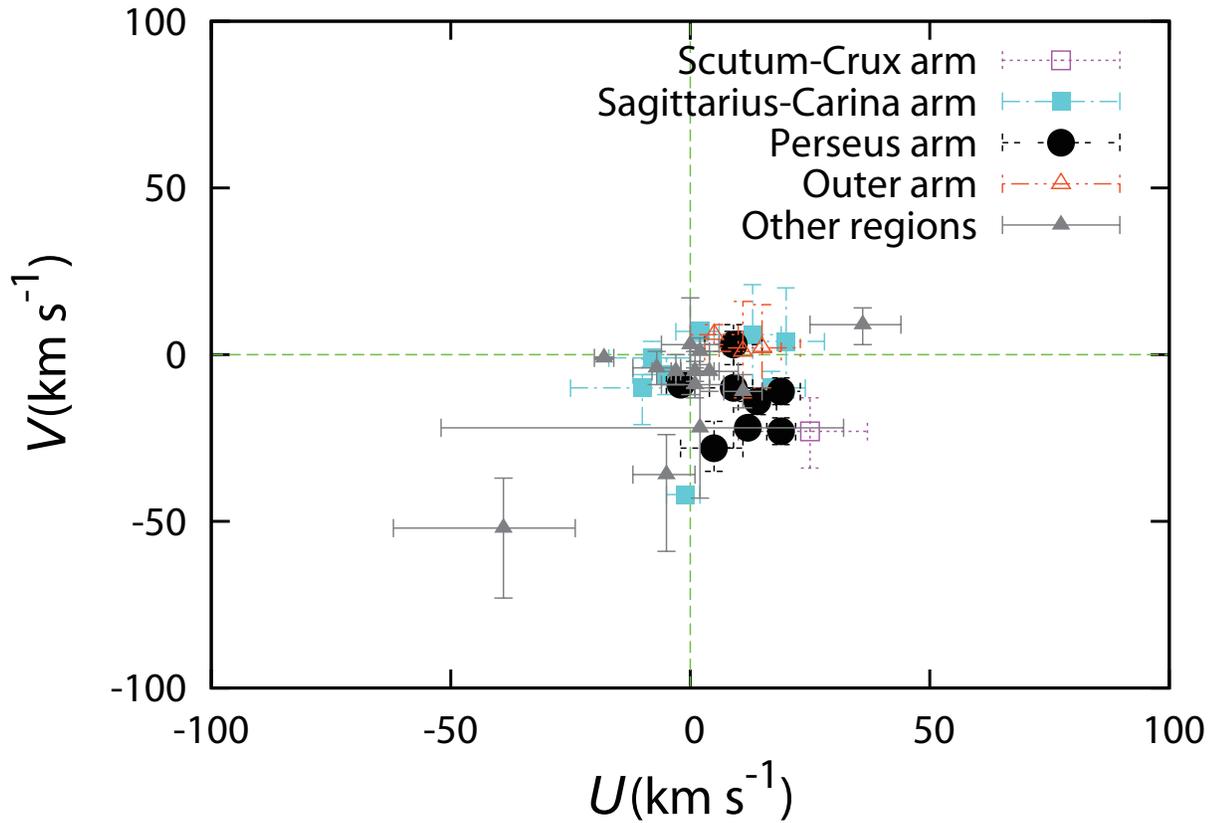}\\
\FigureFile(80.0mm,60.0mm){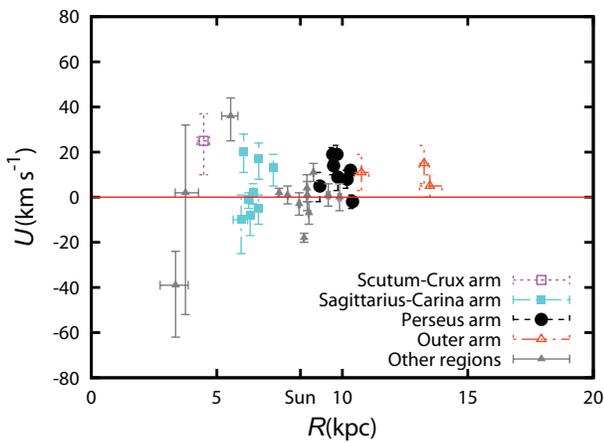}%
\FigureFile(80.0mm,60.0mm){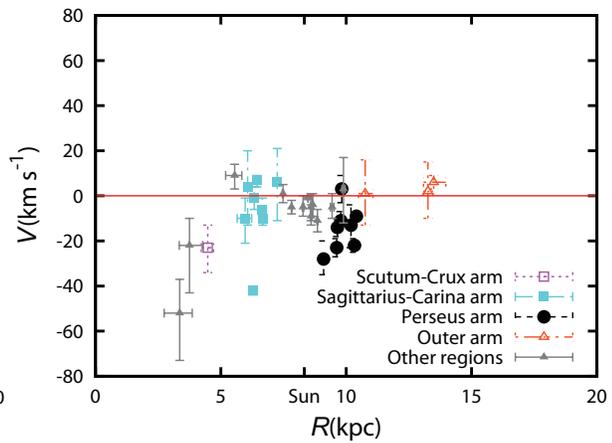}
\end{center}
\caption{(a)Each source of Scutum-Crux (open square), Sagittarius-Carina (filled square), Perseus (circle), Outer arms (open triangle), and other regions (filled triangle) is plotted on peculiar motions plane of $U$ and $V$ for the 33 sources listed in table 6.
 (b)The peculiar motions of $U$ component are shown as a function of the Galacto-centric distance ($R$) for the sources. $R_{0}$ = 8.33 kpc is assumed. $U$ is directed toward the Galactic center. (c)This is the same as (b) but for the peculiar motions of $V$ component. $V$ is directed toward the Galactic rotation.}
\end{figure*}

Next, table 5 shows reduction of physical parameters, virial mass, LTE mass, the ratio of virial mass to LTE mass ($\alpha$), bolometric luminosity, and spectral type due to the distance reduction for IRAS 05168+3634.
Virial mass $M_{\rm{vir}}$ was calculated based on the following equation described by MacLaren et al. (1988),
\begin{equation}
M_{\rm{vir}}=\frac{k_{1}\sigma^{\rm{2}}R}{G},
\end{equation}
where $\sigma$ is the three-dimensional velocity dispersion averaged over the whole system, $R$ is the cloud radius, and $k_{1}$ is the constant whose exact value depends on the form of the density distribution in the cloud.
In contrast,  $M_{\rm{LTE}}$ was also calculated based on both $R$ and peak $^{13}$CO column densities ($M_{\rm{LTE}}$ $\propto$ $R^{\rm{2}}$, Wang et al. 2009). 
The inequality sign (``$>$``) in table 5 shows the lower limit of $M_{\rm{LTE}}$ since Wang et al. (2009) did not widely observe the whole molecular cloud in which IRAS 05168+3634 is located. 
As a result, $\alpha$ = $M_{\rm{vir}}$ / $M_{\rm{LTE}}$ is proportional to $R^{\rm{-1}}$. 
We summarize the ratio $\alpha$ in table 5. 
While the ratio is 0.2 when we adapt the kinematic distance of 6 kpc, the ratio becomes 0.7 ($\sim$6/1.88$\times$0.2) when we adapt our result of 1.88 kpc.
An $\alpha$ close to 1 has been reported (e.g. Wang et al. 2009), 
indicating that most of the disk clouds appear to be virialized. 
Therefore, the latter value 0.7 obtained by adapting our result indicates that the molecular cloud with which IRAS 05168+3634 is associated is also likely to be virialized in the same way as other disk clouds.

\subsection{Rotation velocity of IRAS 05168+3634}\label{sec:math-symbols}
Combination of the distance and the systematic proper motions for IRAS 05168+3634 with the systemic velocity provides full space motion of the source, which allows us to determine not only the rotation velocity of the source, but also its peculiar motions.
First, we converted ($\mu_{\alpha}$cos$\delta$, $\mu_{\delta}$) = (0.23 $\pm$ 1.07,\ $-$3.14 $\pm$ 0.28) mas yr$^{-1}$ in the equatorial coordinate system into ($\mu_{\ell}$cosb, $\mu_{b}$) = (2.71 $\pm$ 0.84,\ $-$1.61 $\pm$ 1.04) mas yr$^{-1}$ in the Galactic coordinate one. 
By adapting the distance of our measurement (d = 1.88$^{+0.21}_{-0.17}$ kpc), we obtain ($\mu_{\ell}$cosb, $\mu_{b}$) = (24.1 $\pm$ 7.5,\ $-$14.3 $\pm$ 9.2) km s$^{-1}$. 
Second, we used $V_{\rm{LSR}}$ = ${-}$15.5 $\pm$ 1.9 km s$^{-1}$ of CS (2-1) emission (Bronfman et al. 1996) as the radial velocity of the source, which is converted into $V_{\rm{helio}}$ = $-$7.8 $\pm$ 1.9 km s$^{-1}$ in the heliocentric flame. 
The conversion calculation was conducted based on the equations in Appendix of Reid et al. (2009b).
 Third, the peculiar solar motions of ($U_{\odot}$, $V_{\odot}$, $W_{\odot}$) = (11.1 $\pm$ 1, 12.24 $\pm$ 2, 7.25 $\pm$ 0.5) km s$^{-1}$ were assumed from Schonrich, Binney, $\&$ Dehnen, (2010).
The peculiar motions represent a deviation of the Sun from the Galactic circular orbit. 
Directions of the peculiar motions are towards the Galactic center ($U_{\odot}$), the Galactic rotation ($V_{\odot}$), and the northern Galactic pole ($W_{\odot}$). 
Fourth, the Galactic constants, $R_{\rm{0}}$ = 8.33 kpc and $\Theta_{0}$ = 240 km s$^{-1}$, were assumed from Reid $\&$ Brunthaler (2004) and Gillessen et al. (2009). 
Reid $\&$ Brunthaler (2004) observed proper motions of Sgr A$^{\ast}$, and Gillessen et al. (2009) determined the distance to the Galactic-center based on stellar orbits around Sgr A$^{\ast}$. 
Finally, these converted and assumed values allow us to determine a rotation velocity ($\Theta$) of 227$^{+9}_{-11}$ km s$^{-1}$ at the source. At the same time, we obtain the peculiar motions of the source in the Galactic plane by simple geometry (e.g. Appendix in Reid et al. 2009b). 
Note that  the  error of rotation velocity is evaluated from errors of the parallax, the proper motions, and the systemic velocity. 
The rotation velocity $\Theta$($R$) is marginally smaller than  the  rotation velocity at the LSR, $\rm{\Theta}_{0}$. 
This result may indicate that the Galactic rotation at the Galacto-centric distance of 10.2 kpc is slower than the rotation velocity at the LSR. 
Note that the source is located at the Galacto-centric distance of 10.19$^{+0.21}_{-0.17}$ kpc. 
The previous six VLBI results in the Perseus arm are consistent with our result of the slower rotation (table 6 and fig. 4). 
The previous VLBI results together with our result are also consistent with the previous one-dimensional (radial velocity) observations called the 9-kpc dip in Sofue et al. (2009). 
In the next section, we will further discuss this slower rotation with the peculiar motions in the Perseus arm.

\subsection{Peculiar motions: Comparison between the Perseus arm and other regions}\label{sec:math-symbols}
 As for the peculiar motions of IRAS 05168+3634, we derive the values of ($U$, $V$, $W$) = (8.1$^{+3.6}_{-3.9}$, $-12$.5$^{+9.2}_{-11.1}$, $-$7.0$^{+9.7}_{-11.9}$) through the procedure described in the previous section. 
Directions of the peculiar motions are towards the Galactic center ($U$), the Galactic rotation ($V$), and the northern Galactic pole ($W$) at the source position. 
The errors of the peculiar motions were evaluated from errors of the parallax, the proper motions, and the systemic velocity.
The peculiar motions tell us a deviation of the source from the circular Galactic orbit. Note that here we assumed the flat rotation model of $\rm{\Theta}$($R$) = $\Theta_{0}$. 
To compare the peculiar motions to previous VLBI results, we list the previous VLBI results in table 6. 
The sources listed in table 6 were observed with VLBI in star-forming regions. 
In figure 4a, we plot the disk peculiar motions on $U$ and $V$ plane from table 6 for the 33 sources. 
Each symbol shows Scutum-Crux (open square), Sagittarius-Carina (filled square), Perseus (circle), Outer arms (open triangle), and other regions (filled triangle). 
It is clear that almost all sources in the Perseus arm are systematically located in the lower right region ($U$ $>$ 0 and $V$ $<$ 0) of the $U$-$V$ plane. 
We emphasize that the sources in the Perseus arm are systematically moving toward the Galactic center ($U$ $>$ 0) and counter to the Galactic rotation ($V$ $<$ 0). 
We obtain disk peculiar motions averaged over the seven sources in the Perseus arm as ($U_{\rm{mean}}$, $V_{\rm{mean}}$) = (11 $\pm$ 3, $-$17 $\pm$ 3) km s$^{-1}$ (NGC 281 excluded). 
Both $U_{\rm{mean}}$ and $V_{\rm{mean}}$ show peculiar motions of greater than 3-$\sigma$ significance in the Perseus arm. 
The peculiar motions in the Perseus arm may trace the streaming motions where the Galactic shock front occurs (Roberts 1969).\\
\ \ \ \ As for other sources, some sources show significantly large peculiar motions (e.g. G 9.62+0.20, G23.01-0.41, and G23.66-0.13). 
All of them are located close to the Galactic bar, which may be affected by the gravitational potential of the central bar. 
Roberts et al. (1979) showed that a bar-like potential can induce strong noncircular motions in the gas flow of $\sim$ 50-150 km s$^{-1}$, 
which is consistent with one of the sources (G 9.62+0.20).
Of course, there are other possibilities for large peculiar motions. For instance, G48.61+0.02 shows a peculiar motion that is counter to the Galactic rotation, larger than 40 km s$^{-1}$. 
It is affected by local phenomena, multiple supernovae (Nagayama et al. 2011b). 
Another interesting feature of the peculiar motions is variation of the peculiar motions as a function of Galacto-centric distance (figures 4b and 4c). 
$V$ values vary among spiral arms in figure 4c. 
Perseus and Norma arms have minus $V$ values with respect to around $V$ = 0 km s$^{-1}$ values or plus values in Outer and Carina-Sagittarius arms.
 In particular, this tendency of Outer and Perseus arms was also suggested by the optical (spectroscopic) observations in Russeil et al. (2007). 
Russeil et al. (2007) argued that this difference between the $V$ components of Outer and Perseus arms may be explained by the streaming motions due to the spiral density-wave.
 These streaming motions produce radial ($U$ component) and azimuthal ($V$ component) residual velocities. 
According to Mel'Nik et al. (1999), the difference in peculiar motions of Outer and Perseus arms may be explained by location of the co-rotation (CR) radius with the density-wave theory.
 Inside the co-rotation radius, radial and azimuthal residual velocities are directed toward the Galactic center and counter to the Galactic rotation, while outside the co-rotation, radial and azimuthal residual velocities are directed away from the Galactic center and toward the Galactic rotation: 
inside and outside the CR, the directions of peculiar motions are the inverse of each other. 
Russeil et al. (2007) determined the co-rotation radius of 12.7 kpc by assuming the co-rotation as the position of $V$ = 0 (see fig. 7 in Russeil et al. 2007). 
This result can explain the $V$ values variation between Outer and Perseus arms in the VLBI observations (fig. 4c). 
On the other hand, $U$ values of both Outer and Perseus arms have the same sign ($U$ $>$ 0), which cannot be explained by the location of the co-rotation set between the two arms (fig. 4b). 
However, the number of Outer arm sources is still small (G75.30+1.32, WB89-437, and S269). 
Thus, more observations of the Outer arm are necessary to confirm whether the interpretation of peculiar motions with the density-wave theory is correct or not.
In contrast, Wada et al. (2011) showed that the spiral features of the gas in the Galactic disk are formed by mechanisms that essentially differ from the Galactic shock in stellar density waves.
 They also showed that, unlike the stream motions in the Galactic shock, the interstellar matter flows into the local potential minima with irregular motions.
Therefore, random irregular motions can be another candidate to explain the observed peculiar motions.

\section{Conclusion}
 Our VLBI observations show results of the parallax and the proper motions for IRAS 05168+3634.
 The parallax is 0.532 $\pm$ 0.053 mas which corresponds to a distance of 1.88$^{+0.21}_{-0.17}$ kpc, and the proper motions are ($\mu_{\alpha}$cos$\delta$, $\mu_{\delta}$) = (0.23 $\pm$ 1.07,\ $-$3.14 $\pm$ 0.28) mas yr$^{-1}$.
The distance is significantly smaller than the previously estimated kinematic distance, being 6 kpc (Molinari et al. 1996). 
According to the drastic change in the source distance, the source is placed in the Perseus arm rather than in the Outer arm. 
The combination of the distance, the proper motions, and the systemic velocity provides rotation velocity of 227$^{+9}_{-11}$ km s$^{-1}$ at the source assuming $\rm{\Theta}_{0}$ = 240 km s$^{-1}$. 
Our result indicates marginally slower rotation with $\sim$ 1-$\sigma$ significance for the Perseus arm, but consistent with the previous VLBI results for six sources in the Perseus arm (NGC 281 excluded).
We also show the disk peculiar motions averaged over the seven sources in the Perseus arm as ($U_{\rm{mean}}$, $V_{\rm{mean}}$) = (11 $\pm$ 3, $-$17 $\pm$ 3) km s$^{-1}$.
Note that here we assume the flat rotation as $\Theta$($R$) = $\Theta_{\rm{0}}$. 
This suggests that the seven sources in the Perseus arm are systematically moving toward the Galactic center, and lag behind the Galactic rotation with more than 3-$\sigma$ significance.
The peculiar motions may be caused in the inner edge of the Perseus arm where a shock front predicted by the density-wave theory occurs (Mel'Nik et al. 1999). 
However, the density wave with CR = 12.7 kpc cannot fully explain the disk peculiar motions between Perseus and Outer arms obtained by VLBI observations.
That they both share the same sign of $U$ values ($U$ $>$ 0) cannot be explained by the density wave, while sign variation of $V$ values between the two arms is consistent with the prediction of the density wave. 
Yet more observations are necessary to confirm whether the density-wave theory can explain the disk peculiar motions of both $U$ and $V$ simultaneously or not over the entire Galactic disk.
 We have been observing H$_{2}$O maser sources located in the star-forming regions to determine a parallax and proper motions with VERA.
Our observations will be used to construct a highly accurate outer rotation curve of the Galaxy. 
This would allow us to determine not only the mass distribution of the Galaxy, but also that of the dark matter.
If characteristic phenomena are seen in individual objects, we will report these together with the outer rotation curve.
\\
\\
 \hspace*{30pt} Finally, we thank VERA project members for the support they offered during observations. We would also like to thank the referee for carefully reading the manuscript.

\begin{table*}
\begin{center}
\caption{Peculiar motions between sources of the Perseus arm and that of other places.\footnotemark[$*$]}
\begin{tabular}{lccccl}

\hline
\hline
Source		&$U$		&$V$			&$W$				&Ref.	&Note\\
		&km s$^{-1}$	&km s$^{-1}$		&km s$^{-1}$					&	&\\
\hline
G 9.62+0.20		&$-39^{+15}_{-23}$		&$-52^{+15}_{-21}$		&$-10^{+6}_{-7}$				&15	&close to 3kpc arm\\
G 12.89+0.49		&$20^{+8}_{-9}$		&$4^{+16}_{-17}$		&$-4^{+9}_{-10}$				&16	&Carina-Sagittarius arm\\
G 14.33-0.64		&$13^{+6}_{-8}$		&$6^{+15}_{-17}$		&$-4^{+15}_{-18}$				&17	&Carina-Sagittarius arm\\
G 15.03-0.68		&$2^{+4}_{-5}$		&$7^{+2}_{-3}$		&$-5^{+2}_{-2}$				&16	&Carina-Sagittarius arm\\
G 23.01-0.41		&$25^{+12}_{-15}$		&$-23^{+10}_{-11}$		&$-1^{+4}_{-5}$				&18	&Crux-Scutum arm\\
G 23.44-0.18		&$2^{+30}_{-54}$		&$-22^{+12}_{-21}$		&$2^{+4}_{-5}$				&18	&Norma arm\\
G 23.66-0.13		&$36^{+8}_{-11}$		&$9^{+5}_{-6}$		&$4^{+1}_{-1}$				&19	&close to the bar\\
G 35.20-0.74		&$-5^{+6}_{-7}$		&$-6^{+5}_{-6}$		&$-8^{+2}_{-3}$				&21	&Carina-Sagittarius arm\\
G 35.20-1.74		&$-10^{+11}_{-15}$		&$-10^{+9}_{-11}$		&$-9^{+4}_{-5}$				&21	&Carina-Sagittarius arm\\
G 48.61+0.02		&$-1^{+3}_{-4}$		&$-42^{+2}_{-2}$		&$6^{+2}_{-2}$				&22	&Carina-Sagittarius arm\\
W51 Main/South	&$-8^{+8}_{-9}$		&$-1^{+5}_{-5}$		&$5^{+6}_{-6}$				&23	&Carina-Sagittarius arm\\
IRAS 19213		&$17^{+7}_{-9}$		&$-10^{+5}_{-3}$		&$-4^{+5}_{-7}$				&1	&Carina-Sagittarius arm\\
G 59.78+0.06		&$2^{+2}_{-2}$		&$1^{+4}_{-4}$		&$-4^{+1}_{-1}$				&25	&Local arm\\
ON1			&$1^{+4}_{-4}$		&$-5^{+3}_{-3}$		&$8^{+4}_{-3}$				&26	&Local arm\\
G 75.30+1.32		&$11^{+8}_{-8}$		&1$^{+15}_{-14}$		&$-18^{+8}_{-9}$				&13	&Outer arm\\

ON2N			&$-3^{+5}_{-5}$		&$-5^{+5}_{-4}$		&$1^{+4}_{-4}$				&27	&\\

AFGL 2789		&$5^{+6}_{-7}$		&$-28^{+8}_{-7}$		&$-11^{+6}_{-8}$				&1	&Perseus arm\\
L 1206			&$1^{+6}_{-7}$		&$-9^{+5}_{-4}$		&$0^{+7}_{-8}$				&2	&Star-froming Region\\
Cep A			&$4^{+6}_{-6}$		&$-5^{+6}_{-6}$		&$-5^{+3}_{-3}$				&3	&Local arm\\
NGC 7538		&$19^{+3}_{-3}$		&$-23^{+4}_{-4}$		&$-10^{+2}_{-2}$				&3	&Perseus arm\\
L 1287			&$11^{+4}_{-4}$		&$-11^{+5}_{-5}$		&$-3^{+3}_{-3}$				&2	&SFR\\
IRAS 00420		&$14^{+4}_{-4}$		&$-14^{+4}_{-4}$		&$-3^{+1}_{-2}$				&4	&SFR in the Perseus arm\\
NGC 281		&$9^{+6}_{-6}$		&$3^{+6}_{-6}$		&$-9^{+3}_{-3}$				&2	&SFR in the Perseus arm\\
W3(OH)			&$19^{+4}_{-4}$		&$-11^{+4}_{-4}$		&$1^{+3}_{-3}$				&7	&Perseus arm\\

WB 89-437		&$15^{+8}_{-9}$		&$2^{+13}_{-12}$		&$1^{+13}_{-14}$				&8	&Outer arm\\

IRAS 05168		&8$^{+4}_{-4}$		&$-$13$^{+9}_{-11}$		&$-$7$^{+10}_{-12}$		&9	&Perseus arm\\
HP-Tau/G2			&$-$18$^{+2}_{-2}$		&$-$1$^{+0.3}_{-0.4}$		&0$^{+1}_{-1}$		&34	&\\
IRAS 06061			&12$^{+2}_{-2}$		&$-$22$^{+3}_{-3}$		&$-$11$^{+2}_{-3}$			&35	&Perseus arm\\

S 252			&$-$2$^{+3}_{-3}$		&$-$9$^{+1}_{-1}$		&$-$2$^{+0.4}_{-0.5}$			&10	&Perseus arm\\
S 255			&0$^{+6}_{-6}$		&3$^{+14}_{-15}$		&3$^{+10}_{-11}$				&2	&SFR\\
S 269			&5$^{+5}_{-5}$		&6$^{+3}_{-3}$		&$-$4$^{+2}_{-3}$				&11	&Outer arm\\
Orion			&$-$7$^{+5}_{-5}$		&$-$4$^{+5}_{-5}$		&3$^{+4}_{-4}$				&12	&\\
G232.6+1.0		&1$^{+5}_{-5}$		&$-$5$^{+6}_{-5}$		&1$^{+3}_{-3}$				&10	&Local arm\\
\hline
\textbf{Sun}			&$U_{\odot}$=11.1$^{+1}_{-1}$		&$V_{\odot}$=12.24$^{+2}_{-2}$			&$W_{\odot}$=7.25$^{+0.5}_{-0.5}$		&\footnotemark[$\dagger$]	&\footnotemark[$\dagger$]Schonrich, Binney, $\&$ Dehnen (2010)	\\
\hline
\multicolumn{4}{@{}l@{}}{\hbox to 0pt{\parbox{180mm}{\footnotesize
\par\noindent \\
\footnotemark[$*$] $\Theta_{0}$=240 km s$^{-1}$ and $R_{0}$=8.33 kpc are assumed to determine the peculiar motions. The rotation model is flat ($\Theta$($R$) = $\Theta_{0}$) assumed.\\
$\bf {References.}$ (1) Oh et al. 2010; (2) Rygl et al. 2010; Sato et al. 2008; (3) Moscadelli et al. 2009; (4) Moellenbrock et al. 2009;\\ (5) Reid et al. 2009b; 
(6) Honma et al. 2011; (7) Xu et al. 2006; (8) Hachisuka et al. 2009; (9) this paper; 
(10) Reid et al. 2009a; \\(11) Honma et al. 2007; (12) Hirota et al. 2007; Menten et al. 2007; (13) Sanna et al. 2012;
(15)Sanna et al. 2009; (16)Xu et al. 2011; \\(17)Sato et al. 2010a; (18)Brunthaler et al. 2009; (19)Bartkiewicz et al. 2008; (20)Dzib et al. 2010; (21)Zhang et al. 2009; 
\\(22)Nagayama et al. 2011b; (23)Sato et al. 2010b; (25)Xu et al. 2009; (26)Nagayama et al. 2011a; (27)Ando et al. 2011;
(34)Torres et al. 2009; \\(35)Niinuma et al. 2011;
}\hss}}

\end{tabular}
\end{center}
\end{table*}

\end{document}